\pdfoutput=1
\documentclass[11pt,twoside,a4paper,cmspaper,final,collab]{cms-tdr}
\def\svnVersion{dc0a462}\def\svnDate{2025/04/08}\def\cmsCernNoTag{CERN-EP-2024-216}\def\cmsCernDate{\today}\def\cmsMessage{Published in Physical Review D as \href{http://dx.doi.org/10.1103/PhysRevD.111.072003}{\doi{10.1103/PhysRevD.111.072003}.}}
\begin{document}\cmsNoteHeader{GEN-22-001}

\newcommand{\lele}  {\ensuremath{\Pell^+\Pell^-}\xspace}
\newcommand{\ptll}  {\ensuremath{\pt(\lele)}\xspace}
\newcommand{\mll}   {\ensuremath{m(\lele)}\xspace}
\newcommand{\pp}    {\ensuremath{\Pp\Pp}\xspace}
\newcommand{\fiq}   {\ensuremath{f_\mathrm{i}(q)}\xspace}
\newcommand{\ufiq}  {\ensuremath{u_{f_\mathrm{i}(q)}}\xspace}
\newcommand{\di}    {\ensuremath{d_\mathrm{i}}\xspace}
\newcommand{\udi}   {\ensuremath{u_{d_\mathrm{i}}}\xspace}
\newcommand{\xlin}  {\ensuremath{\chi^2_{\text{lin.}}}\xspace}
\newcommand{\chisq} {\ensuremath{\chi^2}\xspace}
\newcommand{\qs}    {\ensuremath{q_s}\xspace}
\newcommand{\sqs}   {\ensuremath{\sqrt{s}}\xspace}
\newcommand{\massz} {\ensuremath{m(\PZ)}\xspace}

\newlength\cmsTabSkip\setlength{\cmsTabSkip}{1ex}

\ifthenelse{\boolean{cms@external}}
{\providecommand{\suppMaterialA}{Appendix~\ref{supp:result_table}}}
{\providecommand{\suppMaterialA}{Appendix~\ref{supp:result_table}}}

\providecommand{\figenergyscaling}{\ref{fig:energy-scaling}}
\providecommand{\figisr}{\ref{fig:fit_diffslope-ISRpTmin}}
\providecommand{\figQscaling}{\ref{fig:Q-scaling}}

\ifthenelse{\boolean{cms@external}}
{\providecommand{\figwidth}{0.49}}
{\providecommand{\figwidth}{0.85}}

\ifthenelse{\boolean{cms@external}}
{\providecommand{\figwidtha}{0.49}}
{\providecommand{\figwidtha}{0.6}}

\cmsNoteHeader{GEN-22-001}

\title{Energy-scaling behavior of intrinsic transverse momentum parameters in Drell-Yan simulation}

\date{\today}

\abstract{An analysis is presented based on models of the intrinsic transverse momentum (intrisic \kt) of partons in nucleons by studying the dilepton transverse momentum in Drell--Yan events. Using parameter tuning in event generators and existing data from fixed-target experiments and from hadron colliders, our investigation spans three orders of magnitude in center-of-mass energy and two orders of magnitude in dilepton invariant mass. The results show an energy-scaling behavior of the intrinsic \kt parameters, independent of the dilepton invariant mass at a given center-of-mass energy.}

\hypersetup{
pdfauthor={CMS Collaboration},
pdftitle={Energy-scaling behavior of intrinsic transverse momentum parameters in Drell-Yan simulation},
pdfsubject={CMS},
pdfkeywords={CMS, intrinsic kT, Drell-Yan, nonperturbative QCD}}

\maketitle 
\section{Introduction}
The description of high-energy hadron-hadron collisions relies on the modeling of the nonperturbative regime of quantum chromodynamics (QCD). In particular, the modeling of the intrinsic transverse momenta of partons inside the colliding protons (intrinsic \kt) represents a challenge both experimentally and theoretically. Several approaches exist to describe the intrinsic \kt, including those with transverse-momentum (\pt)-dependent parton distribution functions (PDFs)~\cite{martinez2020transverse,Bubanja:2023nrd,Mendizabal:2023mel}, and those based on first-principle approaches such as lattice QCD~\cite{hagler2010intrinsic,musch2007transverse,musch2011exploring}. The most widely used frameworks for the description of collision events at the LHC are general-purpose Monte Carlo (MC) event generators that include parton shower descriptions, such as \PYTHIA~8~\cite{bierlich2022comprehensive,sjostrand2004multiple}, \HERWIG~7~\cite{bellm2016herwig}, and \textsc{Sherpa}~\cite{bothmann2019event}. They are based on collinear PDFs and usually model the intrinsic \kt as a random variable with a Gaussian distribution, whose width is a tunable parameter. However, the extracted intrinsic \kt parameter does not necessarily correspond to a simple model of nonperturbative physics, but also compensates for deficiencies in the parton shower model. These considerations motivate the study
presented in this paper.

The region of Drell--Yan (DY) production where the lepton-pair system is produced with low \pt, around a few \GeV, is especially suited for the study of the intrinsic \kt models because it provides a clean, high-resolution final state and has been studied widely in collider experiments~\cite{NuSea:2003qoe,Webb:2003bj,Antreasyan:1981eg,PHENIX:2018dwt,D0:1999jba,CDF:1999bpw,D0:2010dbl,CDF:2012brb,CMS:2014dyj,ATLAS:2015iiu,CMS:2021ynu,CMS:2021oex,LHCb:2021huf}. Previous studies illustrate the sensitivity of this region to high-order QCD contributions~\cite{Davies:1984sp} and include their own treatments of nonperturbative
physics effects~\cite{guzzi2014nonperturbative,wei2021exploring,Bacchetta:2017gcc,Scimemi:2017etj,Bertone:2019nxa,Scimemi:2019cmh,Bacchetta:2019sam,Bury:2022czx,Bacchetta:2022awv,Moos:2023yfa,Bacchetta:2024qre,Bacchetta:2018lna}. 

In this paper, we investigate the intrinsic \kt models in \PYTHIA and \HERWIG by parameter tuning to DY measurements, and their implications for the perturbative and nonperturbative-QCD effects in the DY initial-state kinematics. The use of CMS data in this investigation extends the probed energy scale of the hard scattering to 1\TeV.
The experimental data and simulations are described in Section~\ref{sec:data-MC}. The strategy for generator tuning and uncertainty estimation is discussed in Section~\ref{sec:gen-tuning}. Section~\ref{sec:UE-DY} investigates the interdependence between the underlying-event~(UE) and the DY lepton-pair transverse-momentum (\ptll) descriptions. The impacts of the hadron-collision energies and the DY hard-scattering energy scales on the intrinsic \kt tunes are discussed in Section~\ref{sec:energyeffect} and \ref{sec:Qeffect}, respectively. Finally, the results are summarized in Section~\ref{sec:summary}.

\section{Data and simulation}\label{sec:data-MC}
We use DY data measured from various types of hadron collisions at different center-of-mass energies and the corresponding simulations to extract nonperturbative-QCD information and to study the interplay between the intrinsic \kt parameter and the perturbative evolution including the initial-state radiation (ISR). The analysis is performed with the \PYTHIA~8 and \HERWIG~7 event generators, which provide different options of ordering variables and energy thresholds for partonic emissions, and the results are compared with those of the previous studies with \textsc{Cascade}~\cite{Baranov:2021uol,Bubanja:2023nrd,Mendizabal:2023mel}. The UE models of the two generators, which consist of multiple-parton interactions (MPI) and beam remnants (BR), as well as hadronization, ISR and final-state radiation (FSR), have been extensively studied~\cite{cms2015underlying,sirunyan2020extraction,sirunyan2021development}. The measurements of the differential cross section of DY processes as a function of the \ptll are summarized in Table~\ref{tab:runs}, in which the proton-lead ($\Pp\mathrm{Pb}$) measurement data were converted for comparison with \pp collisions by correcting for the number of nucleons, and corresponding simulations are used to tune the intrinsic \kt parameter. Tabulated results are provided in the HEPData record for this analysis~\cite{hepdata}.

\begin{table}[htbp!]
    \centering
    \topcaption{Measurements of the DY differential cross section as a function of \ptll at various center-of-mass energies \sqs from different hadron-collision processes used as inputs for the intrinsic \kt tunes. The \sqs in $\Pp\mathrm{Pb}$ collisions represents the nucleon-nucleon center-of-mass energy. The variable $Q$ represents the energy scale of the hard scattering, approximated by the dilepton invariant mass. The Z boson mass is denoted as \massz.}
    \begin{scotch}{llll}
    Experiment                                  & Collision type        & \sqs [{\GeVns}]   & $Q$ [{\GeVns}]    \\
    \hline
    E866/NuSea~\cite{NuSea:2003qoe,Webb:2003bj} & \pp/pd, fixed target  & 38.8              & 4--12.85          \\
    R209~\cite{Antreasyan:1981eg}               & \pp                   & 62                & 5--8              \\
    PHENIX~\cite{PHENIX:2018dwt}                & \pp                   & 200               & 4.8--8.2          \\
    D0/CDF~\cite{D0:1999jba,CDF:1999bpw}        & $\Pp\Pap$             & 1800              & \massz            \\
    D0/CDF~\cite{D0:2010dbl,CDF:2012brb}        & $\Pp\Pap$             & 1960              & \massz            \\
    CMS~\cite{CMS:2014dyj}                      & \pp                   & 2760              & \massz            \\
    ATLAS~\cite{ATLAS:2015iiu}                  & \pp                   & 8000              & 46--150           \\
    CMS~\cite{CMS:2021ynu}                      & $\Pp\mathrm{Pb}$      & 8160              & 15--120           \\
    CMS~\cite{CMS:2021oex}                      & \pp                   & 13000             & 50--1000          \\
    LHCb~\cite{LHCb:2021huf}                    & \pp                   & 13000             & \massz            
    \end{scotch}
    \label{tab:runs}
\end{table}

Predictions of the DY cross section are produced at next-to-leading order (NLO) using \MGvATNLO~v3.4.1~\cite{alwall2014automated} and the matrix-element computation is matched to the \PYTHIA~8 or \HERWIG~7 generator  in the CMS software interface for parton shower and UE activity modeling. The UE tunes used in this analysis are based on various PDFs, partonic emission orderings, and settings in MPI and BR modelings, as listed below:

\begin{itemize}
    \item \PYTHIA~8 tunes: The \PYTHIA~8.243 generator is used for the \pt-ordered parton shower and hadronization, and the UE model is parameterized by the CP3, CP4, or CP5 tune~\cite{sirunyan2020extraction}. The three tunes have the strong coupling $\alpS(m_{\PZ})$ set to 0.118, use NLO strong-coupling evolution, and employ NLO or next-to-NLO (NNLO) PDFs in the hard scattering, parton showering, ISR, FSR, MPI, and BR modelings.
    \item \HERWIG~7 tunes: The \HERWIG~7.1.4 generator is used for the angular-ordered parton shower and hadronization, and the CH2 and CH3 tunes~\cite{sirunyan2021development} are used for the UE modeling. Both tunes use $\alpS(m_{\PZ}) = 0.118$ with NNLO strong-coupling evolution and the NNLO NNPDF3.1 PDFs for the parton shower, but different \alpS and PDF sets in the MPI and BR modelings.
\end{itemize}

\tolerance=1000 These UE tunes from the CMS Collaboration used measurements at \sqs from 0.9 to 13\TeV by the CDF and CMS experiments, modeling the behavior of UE-sensitive variables across all these energies, as well as the kinematics of high \pt jet, $\ttbar$ and DY production at the CMS.

\tolerance=1000 The intrinsic \kt is modeled similarly in \PYTHIA and \HERWIG by a Gaussian distribution with tunable parameters. In \PYTHIA, the width \qs of the Gaussian distribution is related to the parameter $\textsc{BeamRemnants:primordialkThard}\simeq \qs/\sqrt{2}$, whereas in \HERWIG, it is given by the parameter $\textsc{ShowerHandler:IntrinsicPtGaussian}=\qs$. Either of these parameters is referred to as tunable intrinsic \kt parameter $q$ in the following.\par

\section{Generator tuning strategy}\label{sec:gen-tuning}
In each scenario of \sqs given in Table~\ref{tab:runs}, we assume that the DY differential cross section versus \ptll depends on $q$. Simulated samples of events are generated with different choices of the parameter $q$ sampled from the tuning range, resulting in varied predictions of the DY differential cross section. The dependence of the cross section on $q$ predicted for each bin in \ptll is extracted by interpolating the simulated values, as well as their uncertainty, with polynomial functions \fiq and \ufiq, respectively, in each bin $i$ of \ptll. The \textsc{Professor}~2 software~\cite{buckley2010systematic} is employed to perform the interpolation and compute a goodness of fit (GOF), based on the \chisq distribution, for quantifying the agreement between the interpolated simulated values and those measured in the data. The GOF is measured by the \chisq as a function of $q$\par
\begin{equation}
    \label{eq:GOF}
     \chisq(q)=\sum_i \frac{(\fiq-\di)^2}{\udi^2+\ufiq^2},
\end{equation}
where \di and \udi are the cross section and its uncertainty measured in data in bin $i$, respectively. For measurements performed at $\sqs>1\TeV$, the value of \chisq is obtained by summing over the bins $i$ that correspond to the low-\ptll range 0--10\GeV because of its high sensitivity to the intrinsic \kt model. For measurements at $\sqs<1\TeV$, the whole range of \ptll available in the data, which is below 10\GeV, is considered for computing the value of \chisq. For each measurement at a given \sqs, the final tuned parameter $q$ is taken to be the one that minimizes the respective \chisq.

The uncertainties in the tuned parameter originate from the variations of the \chisq. The uncertainty due to the choice of \ptll range is estimated as the difference between the resulting tuned parameter when considering the \ptll ranges [0, 10] and [0, 15]\GeV for $\sqs>1\TeV$, because the transition from the nonperturbative to perturbative contributions takes place approximately between 10--20\GeV. With regard to the $\sqs<1\TeV$ cases, this uncertainty is estimated as the difference between the parameter tuned using the range [0\GeV, max(\pt)] and [0, 2]\GeV (for $\sqs = 38.8\GeV$) or $[0,\,4]\GeV$ (for $\sqs = 62$ and 200\GeV). The uncertainty due to the choice of the functional form used for the interpolation is taken to be the difference between the parameter tuned using order-5 polynomials as the central value and that using order-3 polynomials as the systematic deviation. The uncertainty from the limited statistical precision in the simulation is estimated by substituting the denominator of Eq.~\eqref{eq:GOF} with $\ufiq^2$ and determining the parameter values that result in a value of \chisq increased by 1 compared to the minimum. The contribution from the uncertainty of the measured data is estimated by repeating the tuning for varied values of the differential DY cross section. These varied values are obtained by sampling from Gaussian distributions centered at the nominal values of measured cross sections and with standard deviations equal to their uncertainties. The covariance between the uncertainties contributed by the measured data for different generator setups is estimated from the covariance between the corresponding tunes obtained from the varied measurements. The other uncertainty sources are approximately uncorrelated among the tunes. The uncertainty from the measurement is dominant in most of the tunes, ranging from 2 to 20\%, depending on the measurement precision.

\section{\texorpdfstring{Underlying-event tunes and DY \ptll}{Underlying-event tunes and DY dilepton transverse momentum}}\label{sec:UE-DY}
\begin{figure}[!ht]
    \centering
\includegraphics[width=\figwidtha\textwidth]{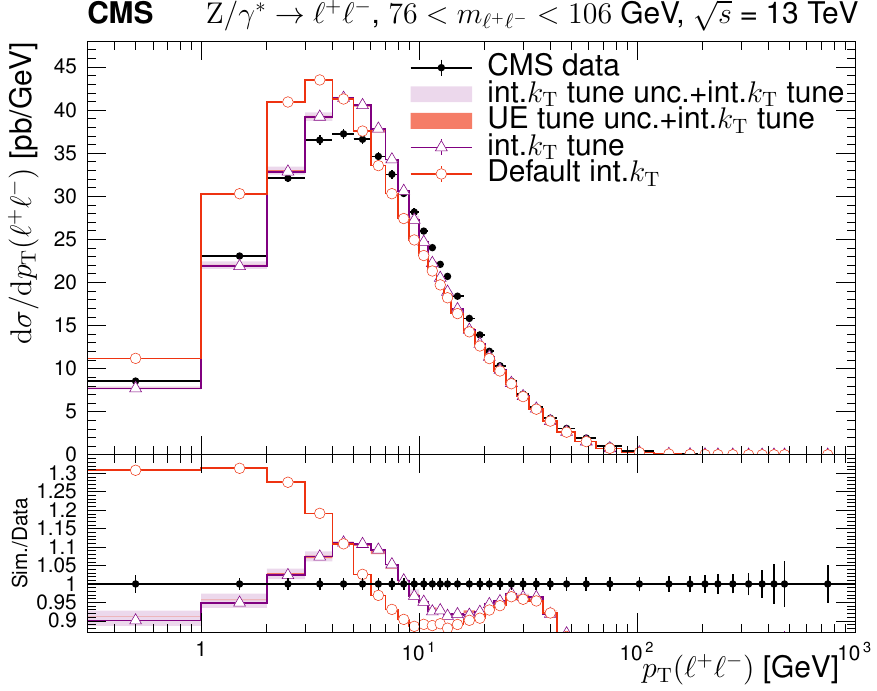}
\includegraphics[width=\figwidtha\textwidth]{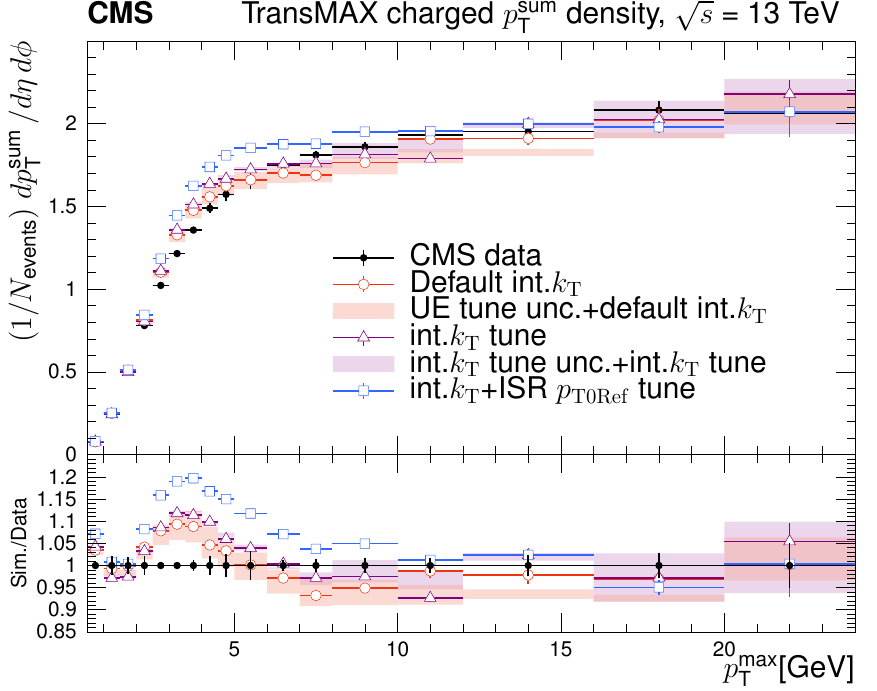}

\caption{Effect of the variation of the UE parameters on the DY \ptll spectrum (upper) and of the variation of the intrinsic \kt parameter on the charged $p_{\mathrm{T}}^{\text{sum}}$ density as a function of $p_{\mathrm{T}}^{\text{max}}$ in the transMAX region of the MB process (lower). The red and violet shaded areas represent the predictions from the up and down variations of the UE tune and the intrinsic \kt tune, respectively. In the upper distribution, both shaded areas are based on the prediction of tuned intrinsic \kt parameter on top of \PYTHIA CP5 (``int.\kt tune''). In the lower distribution, the red shaded areas are based on the prediction of the intrinsic \kt parameter set to the default 1.8 and the UE tune set to \PYTHIA CP5 (``Default int.\kt''), and the violet shaded area is based on the ``int.\kt tune'' prediction. The error bars represent the statistical uncertainty in the simulated events. The lower distribution also includes the UE prediction of the combined tune of the intrinsic \kt and the ISR cutoff scale to the DY \ptll distribution (``int.\kt+ISR $p_{\text{T0Ref}}$ tune''). The data are from the CMS measurements on the DY process~\cite{CMS:2021oex} and the MB process~\cite{cms2015underlying} at 13~\TeV.}

\label{Figure1}
\end{figure}

Our baseline UE description is given by the \PYTHIA CP5 tune~\cite{sirunyan2020extraction}, for which five parameters controlling MPI and color reconnection were varied. To assess the impact of the UE parameterization on the DY dilepton transverse momentum (\ptll) spectrum, we generated DY samples and their \ptll distributions with the intrinsic \kt parameter fixed to the tuned result and the UE parameters set to either the CP5 tune-up or tune-down variation. The difference between the two predictions reflects the effects of the UE-tune uncertainty on the DY \ptll spectrum, shown as red bands in Fig.~\ref{Figure1} (upper), which are much smaller than the uncertainty from the intrinsic \kt variations shown as violet bands.

Similarly, we study the impact of the intrinsic \kt variation on observables that are sensitive to the UE models and have been used to obtain the CP5 tune. An observable used for the UE tuning is the density of the scalar \pt sum of the charged particles ($p_{\mathrm{T}}^{\text{sum}}$) in the pseudorapidity ($\eta$) -- azimuthal angle ($\phi$) space as a function of the \pt of the leading charged particle ($p_{\mathrm{T}}^{\text{max}}$) in the transMAX region averaged over $N_\text{events}$ generated events~\cite{cms2015underlying} in the minimum bias (MB) process. This observable is shown in Fig.~\ref{Figure1} (lower). For each event, the transMAX region is defined by the direction of the leading charged particle in the space transverse to the proton beams. Assuming $\phi$ as the azimuthal angle of the leading charged particle, the ranges of $\phi_1$ satisfying $60^{\circ}<\abs{\phi-\phi_1}<120^{\circ}$ define the two transverse regions, in which transMAX is the one with a higher activity. With the UE parameters fixed to the CP5 tune, we generated MB events with the tuned intrinsic \kt parameter, shown as violet markers in Fig.~\ref{Figure1} (lower). The intrinsic \kt tune uncertainty was estimated from the difference between the tune-up and -down variations, represented by the violet band in Fig.~\ref{Figure1} (lower). Compared with the UE tune uncertainty represented by the red band, the impact of the intrinsic \kt variations on UE-sensitive observables is small. The results shown in Fig.~\ref{Figure1} imply that the parameter space for UE and intrinsic \kt can be factorized. Therefore, in the results presented in the following sections, the intrinsic \kt parameters are tuned without changing the UE tune parameters.

\tolerance=1000 Besides the intrinsic \kt model, the lower cutoff scale of ISR also affects the DY \ptll distribution, which is discussed in Section~\ref{sec:energyeffect}. However, the UE observables are sensitive to the ISR cutoff scale. The combined tune of the intrinsic \kt parameter and the cutoff scale of ISR to the DY \ptll distribution alters the UE observable as shown in Fig.~\ref{Figure1} (lower, blue markers). For the purpose of decoupling the intrinsic \kt study from the UE modeling when investigating its scaling behavior with the collision energy and hard-scattering scale, the parameter for the ISR cutoff is fixed to its default value in the studies shown in Figs.~\ref{fig:energy-scaling} and \ref{fig:Q-scaling}.\par

\section{\texorpdfstring{Impact of the collision energy on the intrinsic \kt tunes}{Impact of the collision energy on the intrinsic transverse momentum tunes}}\label{sec:energyeffect}
\begin{figure}[!ht]
    \centering
    \includegraphics[width=\figwidth\textwidth]{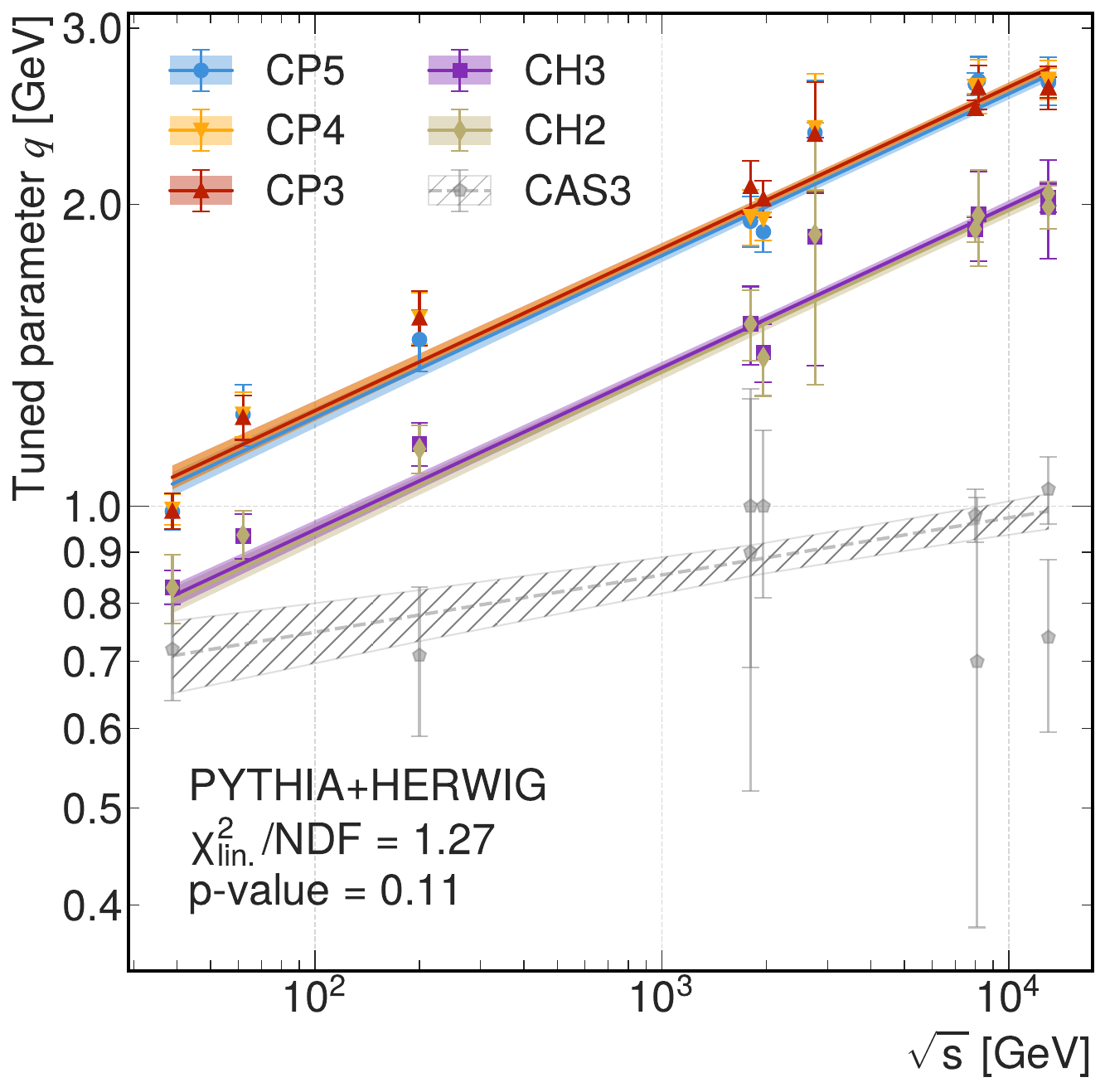}
    \caption{Tuned parameter $q$ values for DY measurements at different center-of-mass energies (points) for various \PYTHIA and \HERWIG setups (colors). The error bars on the points represent the tuning uncertainties. The tuned values are given in \suppMaterialA. For each generator setup, the function $b \sqs^a$ is fitted to the points and shown as a line, assuming the same slope $a$ for all the settings. The $\xlin/\text{NDF}$ and $p$-value of the combined linear fit is given in the plot. The uncertainty in each fit is shown as a colored band and corresponds to the up and down variations of the fit parameters, propagated from the tune uncertainties. The \textsc{Cascade} predictions (CAS3)~\cite{Bubanja:2023nrd,Mendizabal:2023mel} are also fitted separately with the function $b \sqs^a$ for comparison with \PYTHIA and \HERWIG. \label{fig:energy-scaling}
}
\end{figure}

The impact of the center-of-mass energy on the intrinsic \kt parameter is investigated by tuning the generators to the \ptll measurements for each experimental setup in Table~\ref{tab:runs}.
The results show an energy-scaling behavior of the intrinsic \kt parameter for both generators and all setups, as indicated in Fig.~\ref{fig:energy-scaling}. The function $q(\sqs) = b \sqs^a$ is fitted to the points, with fit parameters $a$ and $\log_{10}(b)$ describing the slope and intercept of the function, respectively. The same slope $a$ is assumed for all generator setups based on observation, which is supported by the combined value of \xlin of all linear fits divided by the total number of degrees of freedom (NDF), $\xlin/\text{NDF} = 1.27$, under this assumption and the corresponding $p$-value~\cite{Gross:2010qma} of 0.11. The slope of the fitted function is $a = 0.162 \pm 0.005$. When fitted with free-floating slopes, the resulting slopes are $0.163 \pm 0.006$, $0.164 \pm 0.006$, $0.170 \pm 0.008$, $0.160 \pm 0.008$ and $0.155 \pm 0.007$ for the CP3, CP4, CP5, CH2 and CH3 tunes, respectively, which are compatible with each other. The generators can be improved by implementing the energy scaling of the intrinsic \kt parameters provided by this analysis.

In addition to the intrinsic \kt model, the DY \ptll distribution also receives contributions from ISR. Therefore, the intrinsic \kt parameters in the generators practically compensate the ISR contribution below the cutoff scale in the tunes to DY measured data. To investigate the effect of the ISR cutoff scale on the intrinsic \kt parameter, we perform an energy-dependent intrinsic \kt tune with a different ISR cutoff scale, by varying the regularization parameter \textsc{SpaceShower:pT0Ref} in \PYTHIA and \textsc{SudakovCommon:pTmin} in \HERWIG. The tuning process and determination of the uncertainty follows the same strategy used in the results shown in Fig.~\ref{fig:energy-scaling} in the \PYTHIA CP5 or \HERWIG CH3 setups. The \textsc{SpaceShower:pT0Ref} parameter is changed to 1\GeV from the default value of 2\GeV for \PYTHIA and the \textsc{SudakovCommon:pTmin} parameter is changed to 0.7\GeV from the default value of 1.22\GeV for \HERWIG, such that the simulation can still reasonably describe the observed DY \ptll distribution after the intrinsic \kt tunes. Lowering the cutoff scale induces more low-energy ISR contributions to the low-\ptll DY distribution.

\begin{figure}[!ht]
    \centering
    \includegraphics[width=\figwidth\textwidth]{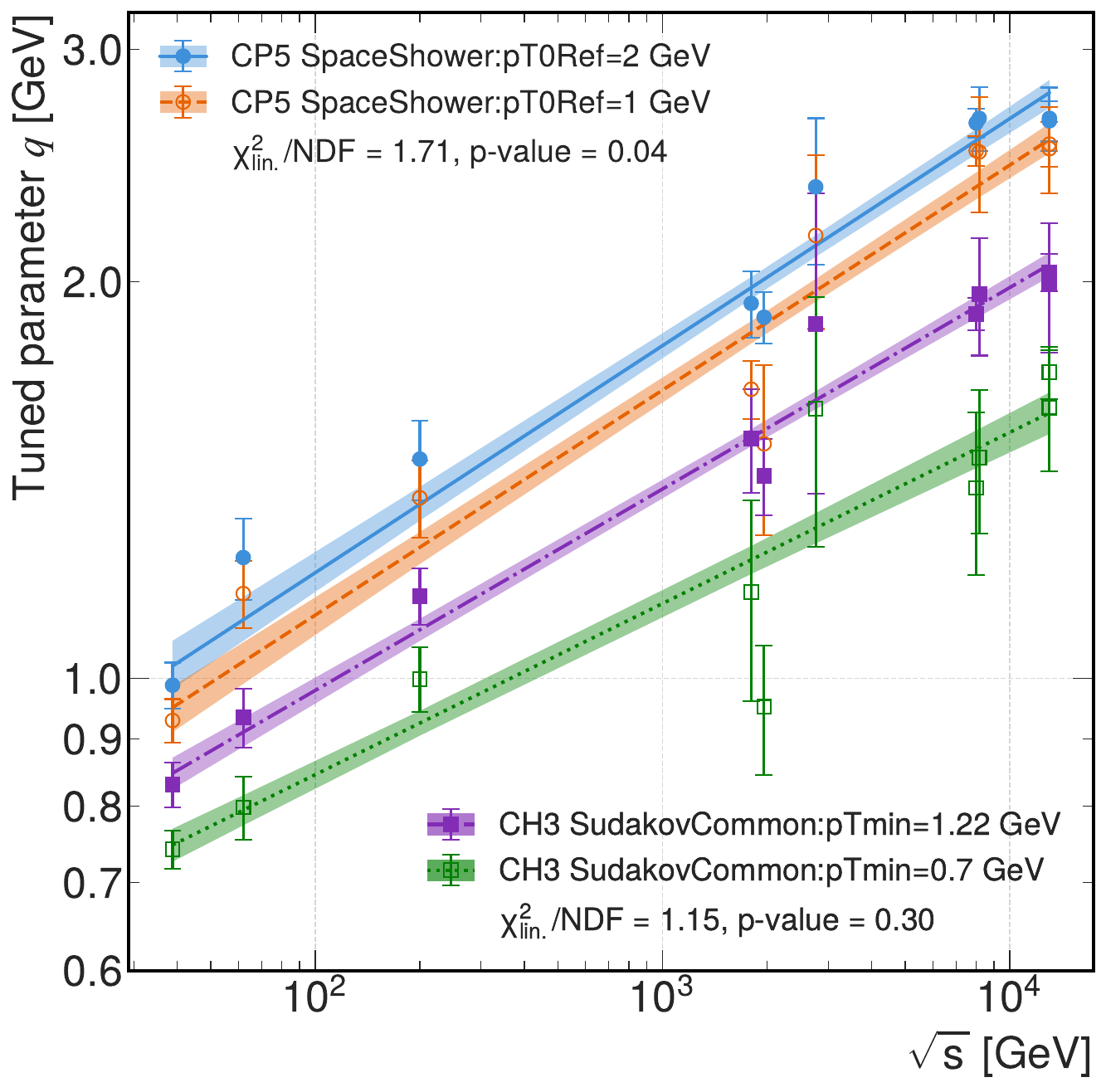}
    \caption{    
        Tuned parameter $q$ values for DY measurements at different center-of-mass energies (points) for various generator settings (lines and bands). The error bars on the points represent the tuning uncertainties. The tuned values are given in \suppMaterialA. For the \PYTHIA CP5 setup, the parameter \textsc{SpaceShower:pT0Ref} is set to 1\GeV (orange dashed) or its default value of 2\GeV (blue solid). For the \HERWIG CH3 setup, the parameter \textsc{SudakovCommon:pTmin} is set to 0.7\GeV (green dotted) or its default value of 1.22\GeV (purple dash-dotted). The function $b \sqs^a$ is fitted to the points of each generator setting and shown as a line, allowing free-floating slopes $a$ and offsets $\log_{10}(b)$. The uncertainty in each fit is shown as a colored band and corresponds to the up and down variations of the fit parameters, propagated from the tune uncertainties. 
    }
    \label{fig:fit_diffslope-ISRpTmin}
\end{figure}

Figure~\ref{fig:fit_diffslope-ISRpTmin} shows the results of the combined fit for the tuned parameter in the \PYTHIA CP5 setup with \textsc{SpaceShower:pT0Ref} set to 1\GeV and the default case and those for the tuned parameter in the \HERWIG CH3 setup with \textsc{SudakovCommon:pTmin} set to 0.7\GeV and the default case. For each generator setting, the function $b\sqs^a$ is fitted to the points and shown as a line, allowing various slopes $a$ and offsets $\log_{10}(b)$. The $\xlin/\text{NDF}$ and $p$ value of the fit are 1.71 and 0.04 for \PYTHIA (1.15 and 0.30 for \HERWIG), respectively. The reported $p$ values are the significance of the $\chi^2$ test with NDF = 16 from 20 tuned results and four parameters to fit, including two slope parameters and two offset parameters. The slope parameters for the \PYTHIA setups are $0.172 \pm 0.009$ (default) and $0.170 \pm 0.009$ (lower ISR cutoff), and the slopes for the \HERWIG setups are $0.153 \pm 0.007$ (default) and $0.130 \pm 0.009$ (lower ISR cutoff). This result shows that the variation of the ISR threshold in the simulation does not change the energy-scaling behavior of intrinsic \kt tunes and causes a mild change of the slope only for very low values of the threshold.

The energy-scaling behavior of the intrinsic \kt tunes can lead to deeper insights about the nonperturbative-QCD contributions in the low-\ptll region of the DY process after extracting and comparing the effects of perturbative QCD in the tunes. On one hand, the tunes are affected by perturbative-QCD models varying among the generator setups because both ISR and intrinsic \kt play a role in describing the \ptll in the DY process. These effects are demonstrated by the variations of the intercepts of linear fit, as shown in the fits in Figs.~\ref{fig:energy-scaling} and~\ref{fig:fit_diffslope-ISRpTmin}. Additionally, the \PYTHIA and \HERWIG generators use different showering models, \PYTHIA using \pt-ordered showering and \HERWIG using angular-ordered showering, which leads to more low-energy ISR in the \HERWIG generation of DY events and less compensation from the nonperturbative part to model \ptll. Therefore, the intercepts in the \HERWIG fit are smaller than those for the \PYTHIA fit in Fig.~\ref{fig:energy-scaling}. On the other hand, the slopes of the linear fits are similar for the various \PYTHIA and \HERWIG tunes in combination with the DY matrix element computed at NLO, despite their differences in the PDF, the order in the parton shower, and the cutoff scale of showering. Stable under variations of perturbative-QCD modeling and hard-scattering scales, this energy-scaling behavior potentially originates from nonperturbative-QCD effects, pointing to the need for further theoretical investigation. The adjustment of intrinsic \kt under higher-order or resummed DY matrix elements and their corresponding energy-scaling behaviors presents a compelling avenue for future research, as nonperturbative-QCD effects are increasingly incorporated into these matrix elements.

A different approach to modeling the intrinsic \kt is implemented in the \textsc{Cascade} generator~\cite{Baranov:2021uol} (``CAS3'' in Fig.~\ref{fig:energy-scaling}),
which accounts for the effect of low-energy gluons with a nonperturbative Sudakov form factor~\cite{wei2021exploring}, using the parton branching method~\cite{martinez2020transverse,Bubanja:2023nrd}. The \textsc{Cascade} generator predicts an intrinsic \kt~\cite{Bubanja:2023nrd,Mendizabal:2023mel} that, compared with \PYTHIA and \HERWIG, is less dependent on the collider energy as shown in Fig.~\ref{fig:energy-scaling}. The energy scaling of the intrinsic \kt parameters in the \PYTHIA and \HERWIG models may be necessary to account for nonperturbative and low-energy gluon emissions not included in the parton showers.

\section{\texorpdfstring{Impact of the hard-scattering scales on the intrinsic \kt tunes}{Impact of the hard-scattering scales on the intrinsic transverse momentum tunes}}\label{sec:Qeffect}

\begin{figure}[!ht]
    \centering
    \includegraphics[width=\figwidth\textwidth]{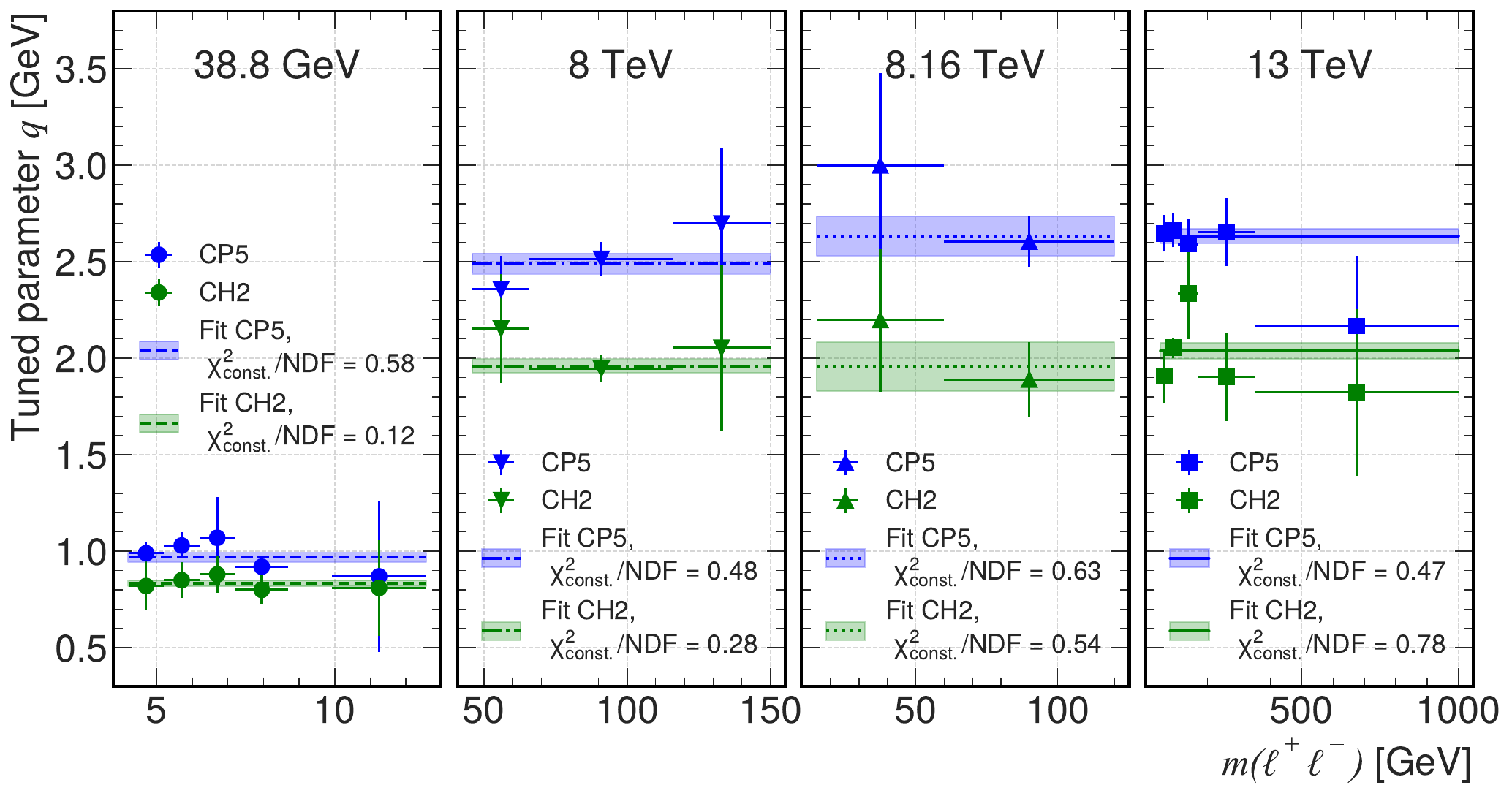}
    \caption{Tuned parameter values (points) for DY measurements at four different center-of-mass energies $\sqs = 38.8\GeV$~\cite{NuSea:2003qoe,Webb:2003bj}, and 8~\cite{ATLAS:2015iiu}, 8.16~\cite{CMS:2021ynu}, and 13\TeV~\cite{CMS:2021oex}, for the \PYTHIA CP5 (blue dark) and \HERWIG CH2 (green light) setups. The error bars on the points represent the tuning uncertainties. The tuned values are given in \suppMaterialA. For each generator setup, a constant is fitted to the points and shown as a line. The uncertainty in each fit, propagated from the tune uncertainties, is shown as a colored band.}
    \label{fig:Q-scaling}
\end{figure}

To explore the dependence of intrinsic \kt parameters on the DY hard-scattering scale $Q$, tunings are performed using the DY differential cross section versus \ptll in exclusive ranges of dilepton invariant mass \mll for $\sqs = 38.8\GeV$~\cite{NuSea:2003qoe,Webb:2003bj}, and 8~\cite{ATLAS:2015iiu}, 8.16~\cite{CMS:2021ynu}, and 13\TeV~\cite{CMS:2021oex}, in which measurements in multiple \mll ranges are available. As shown in Fig.~\ref{fig:Q-scaling}, the intrinsic \kt tunes of either \PYTHIA (CP5) or \HERWIG (CH2)  are similar for various \mll ranges at the same \sqs, which leads to the hypothesis that the intrinsic \kt parameter is identical for all the \mll ranges at a fixed \sqs. Figure~\ref{fig:Q-scaling} shows the results of a constant fit to the tuned parameter values based on this hypothesis, which is compatible with the tuned values as indicated by the value of $\chi^2_{\text{const.}}/\text{NDF}$ being close to 1. This investigation of the $Q$ dependence of the intrinsic \kt tunes under fixed \sqs complements previous studies on the \sqs dependence of the resummed nonperturbative Sudakov factor under fixed $Q$, as summarized in Ref.~\cite{Balazs:1997xd}.

Defining the fractions of the hadron momentum carried by the incoming pair of quarks in the DY process to be $x_1$ and $x_2$, \mll is given by $\mll = x_1 x_2 \sqs$ at leading order. Since the tuned values are stable versus $x_1 x_2 \sqs$ for given values of \sqs, the intrinsic \kt parameter is independent of $x_1 x_2$ within the present precision. Furthermore, the impact of $x_1/x_2$ on the intrinsic \kt tunes is demonstrated by the tunes using the CMS and LHCb measurements at $\sqs = 13\TeV$ because of the different rapidity regions used for the measurements by the two experiments. The tunes to the two measurements agree within the uncertainties (shown in Fig.~\ref{fig:energy-scaling}), which indicates the intrinsic \kt tune to be stable under $x_1/x_2$ variations and suggests its independence of the momentum fractions $x_1$ and $x_2$ individually.

\section{Summary}\label{sec:summary}
In summary, generator tunes of the intrinsic transverse momentum \kt were used to explore model-independent features of nonperturbative quantum chromodynamics (QCD). The tunes were performed for various underlying-event (UE) setups in the generators \PYTHIA and \HERWIG using the Drell--Yan differential cross section as a function of the dilepton transverse momentum measured in multiple types of hadron collision experiments with \sqs ranging from 38.8\GeV to 13\TeV. The results show a linear relation between the logarithm of the intrinsic \kt parameter and $\log_{10}(\sqs)$ for all generator tunes, with the intercepts altered by generator-dependent perturbative-QCD models such as choices of parton distribution functions or parton shower parameters. The slope is $0.162 \pm 0.005$, independent of the UE tune or generator, and related to nonperturbative-QCD effects such as nonresolvable low-energy gluon emissions in parton showers. The tunes were also performed using experiments that probe different regions of dilepton invariant mass and rapidity, and demonstrate the independence of the intrinsic \kt parameter with respect to these variables at each \sqs. This indicates the independence of the intrinsic \kt on the longitudinal momentum fractions of the quarks within colliding hadrons in Drell--Yan processes.

\begin{acknowledgments}
    We congratulate our colleagues in the CERN accelerator departments for the excellent performance of the LHC and thank the technical and administrative staffs at CERN and at other CMS institutes for their contributions to the success of the CMS effort. In addition, we gratefully acknowledge the computing centers and personnel of the Worldwide LHC Computing Grid and other centers for delivering so effectively the computing infrastructure essential to our analyses. Finally, we acknowledge the enduring support for the construction and operation of the LHC, the CMS detector, and the supporting computing infrastructure provided by the following funding agencies: SC (Armenia), BMBWF and FWF (Austria); FNRS and FWO (Belgium); CNPq, CAPES, FAPERJ, FAPERGS, and FAPESP (Brazil); MES and BNSF (Bulgaria); CERN; CAS, MoST, and NSFC (China); MINCIENCIAS (Colombia); MSES and CSF (Croatia); RIF (Cyprus); SENESCYT (Ecuador); ERC PRG, RVTT3 and MoER TK202 (Estonia); Academy of Finland, MEC, and HIP (Finland); CEA and CNRS/IN2P3 (France); SRNSF (Georgia); BMBF, DFG, and HGF (Germany); GSRI (Greece); NKFIH (Hungary); DAE and DST (India); IPM (Iran); SFI (Ireland); INFN (Italy); MSIP and NRF (Republic of Korea); MES (Latvia); LMTLT (Lithuania); MOE and UM (Malaysia); BUAP, CINVESTAV, CONACYT, LNS, SEP, and UASLP-FAI (Mexico); MOS (Montenegro); MBIE (New Zealand); PAEC (Pakistan); MES and NSC (Poland); FCT (Portugal); MESTD (Serbia); MCIN/AEI and PCTI (Spain); MOSTR (Sri Lanka); Swiss Funding Agencies (Switzerland); MST (Taipei); MHESI and NSTDA (Thailand); TUBITAK and TENMAK (Turkey); NASU (Ukraine); STFC (United Kingdom); DOE and NSF (USA).
\end{acknowledgments}

\bibliography{auto_generated}

\appendix
\numberwithin{table}{section}
\numberwithin{figure}{section}
\section{Tuning results} \label{supp:result_table}
Table~\ref{tab:tune-energy-allunc} gives the tuning results shown in Fig.~\figenergyscaling. Tables~\ref{tab:tune-energy-ISRpT0Ref} and~\ref{tab:tune-energy-ISRpTminp7} give the tuning results corresponding to the entries ``CP5 SpaceShower:pT0Ref = 1\GeV'' and ``CH3 SudakovCommon:pTmin = 0.7\GeV'' in Fig.~\figisr, respectively. Table~\ref{tab:tune-dilepton-mass} gives the tuning results shown in Fig.~\figQscaling.

\begin{table*}[htb]
    \centering
    \topcaption{Tune results for the \textsc{BeamRemnants:primordialkThard} parameter in \PYTHIA~8 and the \textsc{ShowerHandler:IntrinsicPtGaussian} parameter in \HERWIG7, taking into account the statistical uncertainty of simulations (MC stat.), the measurement uncertainty in data (data unc.), the uncertainty from choices of tune ranges (range) and the uncerainty from choises of the functions for interpolation (int.).}
    \renewcommand{\arraystretch}{0.91}
    \begin{scotch}{ccc}
       \sqs & Generator setup & Tune result $\pm$ MC stat. $\pm$ data unc. $\pm$ range $\pm$ int. \\
       \hline
        38.8\GeV & \PYTHIA8 CP5 & $0.988 \pm 0.0008 \pm 0.029 \pm 0.022 \pm 0.015$ \\
        & \PYTHIA8 CP4 & $0.993 \pm 0.0008 \pm 0.029 \pm 0.017 \pm 0.009 $ \\
        &\PYTHIA8 CP3 & $0.990 \pm  0.004 \pm 0.03 \pm 0.017 \pm 0.020 $ \\
        & \HERWIG7 CH2 & $0.829 \pm 0.0005 \pm 0.017 \pm 0.010 \pm 0.06$ \\
        & \HERWIG7 CH3 & $0.830 \pm 0.0005 \pm 0.017 \pm 0.010 \pm 0.026$ \\
        [\cmsTabSkip]
        62\GeV &\PYTHIA8 CP5 & $1.24 \pm 0.0008 \pm 0.07 \pm 0.0015 \pm 0.06  $ \\
        &\PYTHIA8 CP4 & $1.24 \pm 4 \times 10^{-8}  \pm 0.06 \pm 0.0012 \pm 0.006  $ \\
        &\PYTHIA8 CP3 & $1.23 \pm 0.0009 \pm 0.06 \pm 0.0010 \pm 0.012 $ \\
        &\HERWIG7 CH2 & $0.94 \pm 0.0006 \pm 0.05 \pm 0.0012 \pm 0.024 $ \\
        &\HERWIG7 CH3 & $0.93 \pm 0.0006 \pm 0.04 \pm 0.0014 \pm 0.019$ \\
        [\cmsTabSkip]
        200\GeV &\PYTHIA8 CP5 & $1.47 \pm 0.0022 \pm 0.08 \pm 0.005 \pm 0.06$ \\
        &\PYTHIA8 CP4 & $1.54 \pm 0.0024 \pm 0.09 \pm 0.003 \pm 0.004$ \\
        &\PYTHIA8 CP3 & $1.54 \pm 0.0024 \pm 0.09 \pm 0.003 \pm 0.022$ \\
        &\HERWIG7 CH2 & $1.14 \pm 0.0014 \pm 0.06 \pm 0.003 \pm 0.018$ \\
        &\HERWIG7 CH3 & $1.15 \pm 0.0015 \pm 0.06 \pm 0.005 \pm 0.004$ \\
        [\cmsTabSkip]
        1.8\TeV & \PYTHIA8 CP5 & $1.93 \pm 0.012 \pm 0.11  \pm 0.02  \pm 0.015$ \\
        & \PYTHIA8 CP4 & $1.94 \pm 0.013 \pm 0.12 \pm 0.04  \pm 0.0005 $ \\
        &\PYTHIA8 CP3 & $2.09 \pm 0.013 \pm 0.12 \pm 0.03  \pm 0.007 $ \\
        & \HERWIG7 CH2 & $1.52 \pm 0.014 \pm 0.12 \pm 0.03  \pm 0.0024$ \\
        & \HERWIG7 CH3 & $1.52 \pm 0.014 \pm 0.13 \pm 0.03 \pm 0.010$ \\
        [\cmsTabSkip]
        1.96\TeV & \PYTHIA8 CP5 & $1.88 \pm 0.017 \pm 0.08  \pm 0.10  \pm 0.009 $ \\
        &\PYTHIA8 CP4 & $1.93 \pm 0.018 \pm 0.08  \pm 0.03  \pm 0.009 $ \\
        & \PYTHIA8 CP3 & $2.03 \pm 0.019 \pm 0.08  \pm 0.03  \pm 0.008 $ \\
        & \HERWIG7 CH2 & $1.41 \pm 0.023 \pm 0.10  \pm 0.08  \pm 0.0019$ \\
        & \HERWIG7 CH3 & $1.42 \pm 0.021 \pm 0.09 \pm 0.024 \pm 0.021$ \\
        [\cmsTabSkip]
        2.76\TeV & \PYTHIA8 CP5 & $2.36 \pm 0.022 \pm 0.3  \pm 0.005  \pm 0.005$ \\
        & \PYTHIA8 CP4 & $2.39 \pm 0.020 \pm 0.3  \pm 0.024  \pm 0.013$ \\
        & \PYTHIA8 CP3 & $2.35 \pm 0.023 \pm 0.3  \pm 0.004  \pm 0.007$ \\
        & \HERWIG7 CH2 & $1.87 \pm 0.03 \pm 0.5  \pm 0.06  \pm 0.003$ \\
        & \HERWIG7 CH3 & $1.9 \pm 0.03 \pm 0.4 \pm 0.18 \pm 0.018$ \\
        [\cmsTabSkip]
        8\TeV & \PYTHIA8 CP5 & $2.64 \pm 0.006 \pm 0.0251  \pm 0.06  \pm 0.0016$ \\
        &\PYTHIA8 CP4 & $2.62 \pm 0.008 \pm 0.022  \pm 0.04  \pm 0.006$ \\
        & \PYTHIA8 CP3 & $2.50 \pm 0.008 \pm 0.012  \pm 0.03  \pm 0.026$ \\
        & \HERWIG7 CH2 & $1.89\pm 0.008 \pm 0.015  \pm 0.05  \pm 0.007 $ \\
        & \HERWIG7 CH3 & $1.89 \pm 0.007 \pm 0.009 \pm 0.05 \pm 0.007$ \\
        [\cmsTabSkip]
        8.16\TeV & \PYTHIA8 CP5 & $2.66 \pm 0.029 \pm 0.14  \pm 0.02  \pm 0.015$ \\
        & \PYTHIA8 CP4 & $2.63 \pm 0.029 \pm 0.16  \pm 0.023  \pm 0.013$ \\
        & \PYTHIA8 CP3 & $2.62 \pm 0.029 \pm 0.13  \pm 0.015  \pm 0.007$ \\
        & \HERWIG7 CH2 & $1.96 \pm 0.03 \pm 0.19  \pm 0.06 \pm 0.027$ \\
        & \HERWIG7 CH3 & $1.96 \pm 0.03 \pm 0.17 \pm 0.09 \pm 0.02$ \\
        [\cmsTabSkip]
        13\TeV & \PYTHIA8 CP5 & $2.648 \pm 0.006 \pm 0.027  \pm 0.028  \pm 0.04$ \\
        (CMS)& \PYTHIA8 CP4 & $2.654 \pm 0.006 \pm 0.027  \pm 0.08  \pm 0.004$ \\
        & \PYTHIA8 CP3 & $2.619 \pm 0.006 \pm 0.028 \pm 0.05  \pm 0.009$ \\
        & \HERWIG7 CH2 & $2.05 \pm 0.03 \pm 0.04  \pm 0.03  \pm 0.035$ \\
        & \HERWIG7 CH3 & $2.03 \pm 0.03 \pm 0.04 \pm 0.021 \pm 0.010$ \\
        [\cmsTabSkip]
        13\TeV & \PYTHIA8 CP5 & $2.66 \pm 0.009 \pm 0.06  \pm 0.13  \pm 0.0007$ \\
        (LHCb)& \PYTHIA8 CP4 & $2.67 \pm 0.009 \pm 0.06  \pm 0.10  \pm 0.003$ \\
        & \PYTHIA8 CP3 & $2.62 \pm 0.009 \pm 0.06  \pm 0.11  \pm 0.00007$ \\
        & \HERWIG7 CH2 & $1.99 \pm 0.04 \pm 0.06 \pm 0.05  \pm 0.04$ \\
        & \HERWIG7 CH3 & $1.99 \pm 0.06 \pm 0.14\pm 0.16 \pm 0.06$ \\
    \end{scotch}
    \label{tab:tune-energy-allunc}
\end{table*}

\begin{table*}[htb]
    \centering
    \topcaption{Tune results for the \textsc{BeamRemnants:primordialkThard} parameter in \PYTHIA8 with the CP5 tune setup, taking into account the statistical uncertainty of simulations (MC stat.), the measurement uncertainty in data (data unc.), the uncertainty from choices of tune ranges (range) and the uncerainty from choises of the functions for interpolation (int.). The parameter \textsc{SpaceShower:pT0Ref} was set to 1\GeV.}
    \begin{scotch}{cc}
       \sqs & Tune result $\pm$ MC stat. $\pm$ data unc. $\pm$ range $\pm$ int.\\
       \hline
       38.8\GeV & $0.929 \pm 0.001 \pm 0.03  \pm 0.015 \pm 0.0005$\\
       62\GeV & $1.16 \pm 1.8 \times 10^{-10} \pm 0.07 \pm 0.0014\pm 0.00018$\\
       200\GeV & $1.37 \pm 0.003 \pm 0.09 \pm 0.006 \pm 0.003$\\
       1.8\TeV & $1.66 \pm 0.013 \pm 0.08 \pm 0.007 \pm 0.016$\\
       1.96\TeV & $1.51 \pm 0.016 \pm0.11 \pm 0.18 \pm 0.08$\\
       2.76\TeV & $2.2 \pm 0.026 \pm 0.3 \pm 0.006 \pm 0.025$\\
       8\TeV & $2.51 \pm 0.04 \pm 0.03 \pm 0.008 \pm 0.04$\\
       8.16\TeV & $2.51 \pm 0.05 \pm 0.20 \pm 0.021 \pm 0.14$\\
       13\TeV (CMS) & $2.54 \pm 0.008 \pm 0.04 \pm 0.09 \pm 0.0024$\\
       13\TeV (LHCb) & $2.52 \pm 0.014 \pm 0.09 \pm 0.17 \pm 0.0021$\\
    \end{scotch}
    \label{tab:tune-energy-ISRpT0Ref}
\end{table*}

\begin{table*}[htb]
    \centering
    \topcaption{Tune results for the \textsc{ShowerHandler:IntrinsicPtGaussian} parameter in \HERWIG7 with the CH3 tune setup, taking into account the statistical uncertainty of simulations (MC stat.), the measurement uncertainty in data (data unc.), the uncertainty from choices of tune ranges (range) and the uncerainty from choises of the functions for interpolation (int.). The parameter \textsc{SudakovCommon:pTmin} was set to 0.7\GeV.}
    \label{tab:tune-energy-ISRpTminp7}
    \begin{scotch}{cc}
       \sqs & Tune result $\pm$ MC stat. $\pm$ data unc. $\pm$ range $\pm$ int.\\
       \hline
       38.8\GeV & $0.742 \pm 0.0010 \pm 0.024 \pm 0.0010 \pm 0.004 $\\
       62\GeV & $0.80 \pm 0.00021 \pm 0.04 \pm 0.0020 \pm 0.00024$\\
       200\GeV & $1.00 \pm 0.0018 \pm 0.06 \pm 0.004 \pm 0.0023$\\
       1.8\TeV & $1.16 \pm 0.024 \pm 0.18 \pm 0.06 \pm 0.07$\\
       1.96\TeV & $0.95 \pm 0.025 \pm 0.10 \pm 0.012 \pm 0.005$\\
       2.76\TeV & $1.60\pm 0.018 \pm 0.23 \pm 0.10 \pm 0.24$\\
       8\TeV & $1.395 \pm 0.07 \pm 0.17 \pm 0.14 \pm 0.08$\\
       8.16\TeV & $1.47 \pm 0.028 \pm 0.17 \pm 0.03 \pm 0.04$\\
       13\TeV (CMS) & $1.71 \pm 0.025 \pm 0.016 \pm 0.00017 \pm 0.07$\\
       13\TeV (LHCb) & $1.60 \pm 0.06 \pm 0.08 \pm 0.13 \pm 0.05$\\
    \end{scotch}
\end{table*}

\begin{table*}[htb]
     \centering
    \topcaption{Results of the tune to various ranges of the \mll for values of \sqs of 38.8\GeV and 8, 8.16, and 13\TeV, taking into account the statistical uncertainty of simulations (MC stat.), the measurement uncertainty in data (data unc.), the uncertainty from choices of tune ranges (range) and the uncerainty from choises of the functions for interpolation (int.).}
     \begin{scotch}{cccc}
       \sqs  & \mll range & \multicolumn{2}{c}{Tune result $\pm$ MC stat. and data unc. $\pm$ range $\pm$ int.}\\
    \hline
    \multicolumn{2}{c}{} & \PYTHIA CP5 &  \HERWIG CH2 \\
     \hline
        38.8\GeV  & 4.2 -- 5.2\GeV & $0.99 \pm 0.05 \pm 0.020 \pm 0.010$ & $0.82 \pm 0.03 \pm 0.011 \pm 0.05$\\
                  & 5.2 -- 6.2\GeV & $1.03 \pm 0.06 \pm 0.020 \pm 0.025$ & $0.85 \pm 0.03 \pm 0.010 \pm 0.09$\\
                  & 6.2 -- 7.2\GeV & $1.07 \pm 0.08 \pm 0.010 \pm 0.20$ & $0.88 \pm 0.05 \pm 0.010 \pm 0.11$\\
                  & 7.2 -- 8.7\GeV & $0.92 \pm 0.04 \pm 0.025 \pm 0.005$ &$ 0.80 \pm 0.03 \pm 0.016 \pm 0.05$\\
                  & 10.2 -- 12.85\GeV & $0.87 \pm 0.31 \pm 0.18 \pm 0.16$ & $0.81 \pm 0.23 \pm 0.09 \pm 0.06$\\
        [\cmsTabSkip]
        8\TeV & 46 -- 66\GeV & $2.36 \pm 0.17 \pm 0.0016 \pm 0.05$ & $2.15 \pm 0.27 \pm 0.03 \pm 0.07$\\
              & 66 -- 116\GeV & $2.51 \pm 0.07 \pm 0.017 \pm 0.05$ & $1.95 \pm 0.05 \pm 0.04 \pm 0.011$\\
              & 116 -- 150\GeV & $2.70 \pm 0.33 \pm 0.13 \pm 0.17$ & $2.1 \pm 0.4 \pm 0.0005 \pm 0.006$\\
        [\cmsTabSkip]
        8.16\TeV & 15 -- 60\GeV & $3.0 \pm 0.4 \pm 0.19 \pm 0.10$ & $2.2 \pm 0.3 \pm 0.10 \pm 0.11$\\
                 & 60 -- 120\GeV & $2.61 \pm 0.13 \pm 0.033 \pm 0.009$ & $1.89 \pm 0.18 \pm 0.08 \pm 0.003$\\
        [\cmsTabSkip]
        13\TeV & 50 -- 76\GeV & $2.65 \pm 0.07 \pm 0.06 \pm 0.017$ & $1.91 \pm 0.14 \pm 0.02 \pm 0.007$\\
               & 76 -- 106\GeV & $2.66 \pm 0.03 \pm 0.08 \pm 0.003$ & $2.05 \pm 0.05 \pm 0.02 \pm 0.01$\\
               & 106 -- 170\GeV & $2.59 \pm 0.07 \pm 0.11 \pm 0.03$ & $2.34 \pm 0.23 \pm 0.07 \pm 0.16$\\
               & 170 -- 350\GeV & $2.65 \pm 0.16 \pm 0.07 \pm 0.007$ & $1.90 \pm 0.16 \pm 0.16 \pm 0.02$\\
               & 350 -- 1000\GeV & $2.17 \pm 0.4 \pm 0.018 \pm 0.017$ & $1.8 \pm 0.4 \pm 0.13 \pm 0.03$\\
     \end{scotch}
     \label{tab:tune-dilepton-mass}
 \end{table*}

\cleardoublepage \section{The CMS Collaboration \label{app:collab}}\begin{sloppypar}\hyphenpenalty=5000\widowpenalty=500\clubpenalty=5000
\cmsinstitute{Yerevan Physics Institute, Yerevan, Armenia}
{\tolerance=6000
A.~Hayrapetyan, A.~Tumasyan\cmsAuthorMark{1}\cmsorcid{0009-0000-0684-6742}
\par}
\cmsinstitute{Institut f\"{u}r Hochenergiephysik, Vienna, Austria}
{\tolerance=6000
W.~Adam\cmsorcid{0000-0001-9099-4341}, J.W.~Andrejkovic, T.~Bergauer\cmsorcid{0000-0002-5786-0293}, S.~Chatterjee\cmsorcid{0000-0003-2660-0349}, K.~Damanakis\cmsorcid{0000-0001-5389-2872}, M.~Dragicevic\cmsorcid{0000-0003-1967-6783}, P.S.~Hussain\cmsorcid{0000-0002-4825-5278}, M.~Jeitler\cmsAuthorMark{2}\cmsorcid{0000-0002-5141-9560}, N.~Krammer\cmsorcid{0000-0002-0548-0985}, A.~Li\cmsorcid{0000-0002-4547-116X}, D.~Liko\cmsorcid{0000-0002-3380-473X}, I.~Mikulec\cmsorcid{0000-0003-0385-2746}, J.~Schieck\cmsAuthorMark{2}\cmsorcid{0000-0002-1058-8093}, R.~Sch\"{o}fbeck\cmsorcid{0000-0002-2332-8784}, D.~Schwarz\cmsorcid{0000-0002-3821-7331}, M.~Sonawane\cmsorcid{0000-0003-0510-7010}, W.~Waltenberger\cmsorcid{0000-0002-6215-7228}, C.-E.~Wulz\cmsAuthorMark{2}\cmsorcid{0000-0001-9226-5812}
% ...
\par}
\vskip\cmsinstskip
\dag:~Deceased\\
$^{1}$Also at Yerevan State University, Yerevan, Armenia\\
$^{2}$Also at TU Wien, Vienna, Austria\\
% ...
\end{sloppypar}
%%% END EDITABLE REGION %%%
% skeleton_end
\end{document}